\documentclass[12pt]{article}

\usepackage{graphics,graphpap, calc}

\newcounter{hours}\newcounter{minutes}
\newcommand{\mydate}{%
\setcounter{hours}{\time/60}%
\setcounter{minutes}{\time-\value{hours}*60}%
\ifnum\theminutes<10
 \today, \thehours:0\theminutes
\else 
 \today, \thehours:\theminutes 
\fi}

\newcommand{\bea}{\begin{eqnarray}}
\newcommand{\eea}{\end{eqnarray}}
\newcommand{\beq}{\begin{equation}}
\newcommand{\eeq}{\end{equation}}

\newcommand\MSbar{{\overline{\rm MS}}}

\renewcommand\sf{\ifmmode{\tilde{f}} \else{$\tilde{f}$} \fi}
\newcommand\st{\ifmmode{\tilde{t}} \else{$\tilde{t}$} \fi}
\renewcommand\sb{\ifmmode{\tilde{b}} \else{$\tilde{b}$} \fi}
\newcommand\sq{\ifmmode{\tilde{q}} \else{$\tilde{q}$} \fi}
\newcommand\sg{\ifmmode{\tilde{g}} \else{$\tilde{g}$} \fi}
\newcommand\bbar{\ifmmode{\bar{b}} \else{$\bar{b}$} \fi}
\newcommand\tbar{\ifmmode{\bar{t}} \else{$\bar{t}$} \fi}
\newcommand\qbar{\ifmmode{\bar{q}} \else{$\bar{q}$} \fi}

\newcommand{\gsim}{\;\raisebox{-0.9ex}
           {$\textstyle\stackrel{\textstyle >}{\sim}$}\;}

\newcommand\sW{\sin\theta_W}
\newcommand\cW{\cos\theta_W}

\newcommand{\LL}{{\cal L}}

\renewcommand{\Re}{{\, {\rm Re}}}

\newcommand\cM{\ifmmode{{\mathcal{M}}} \else{${\mathcal{M}}$} \fi}
\newcommand\cMt[1]{{\mathcal{M}}^{#1}_{\mathrm{tree}}}
\newcommand\cMl[1]{{\mathcal{M}}^{#1}_{\mathrm{loop}}}
\newcommand\ACP{\ifmmode{A_{\mathrm{CP}}} \else{$A_{\mathrm{CP}}$} \fi}

\newcommand\LpL{\Lambda_{(+)}^{L}}
\newcommand\LmL{\Lambda_{(-)}^{L}}
\newcommand\LpR{\Lambda_{(+)}^{R}}
\newcommand\LmR{\Lambda_{(-)}^{R}}
\newcommand\PpL{\Pi_{(+)}^{L}}
\newcommand\PmL{\Pi_{(-)}^{L}}
\newcommand\PpR{\Pi_{(+)}^{R}}
\newcommand\PmR{\Pi_{(-)}^{R}}

\newcommand\Wm[1]{W^{#1}_{\mu}}

\newcommand{\bino}{\tilde{b}}      
\newcommand\wt[1]{\tilde{\omega}{}^{#1}_{}}
\newcommand\charp[1]{\chi{}^{+}_{#1}}
\newcommand\neutr[1]{\chi{}^{0}_{#1}}
\newcommand\higgsinofield[2]{\tilde{h}{}^{#2}_{#1}}

\newcommand\squark[2]{\tilde{#1}{}^{}_{#2}}
\newcommand\Charp[1]{\tilde{\chi}{}^{+}_{#1}}
\newcommand\Charm[1]{\tilde{\chi}{}^{-}_{#1}}
\newcommand\Neutr[1]{\tilde{\chi}{}^{0}_{#1}}

\newcommand\squarkq[2]{\tilde{#1}{}^{*}_{#2}}
\newcommand\Charpq[1]{\bar{\tilde{\chi}}{}^{+}_{#1}}

\newcommand\Neutrq[1]{\bar{\tilde{\chi}}{}^{0}_{#1}}

\newcommand\charge[1]{e_{#1}}
\newcommand\isospin[1]{I^{#1}_{3L}}

\newcommand\massmatrix[2]{{\mathcal M}^{#1}_{#2}}
\newcommand\massQmatrix[2]{({\mathcal M}^{2}_{})^{#1}_{#2}}

\newcommand\Ocn[2]{O^{#1}_{#2}}

\newcommand\anL[2]{b^{\tilde{#1}}_{#2}}
\newcommand\anR[2]{a^{\tilde{#1}}_{#2}}
\newcommand\bcL[2]{k^{\tilde{#1}}_{#2}}
\newcommand\bcR[2]{l^{\tilde{#1}}_{#2}}

\newcommand\anLq[2]{b^{\tilde{#1} *}_{#2}}
\newcommand\anRq[2]{a^{\tilde{#1} *}_{#2}}
\newcommand\bcLq[2]{k^{\tilde{#1} *}_{#2}}
\newcommand\bcRq[2]{l^{\tilde{#1} *}_{#2}}

\newcommand\fL[2]{f^{\tilde{#1}}_{L#2}}
\newcommand\fR[2]{f^{\tilde{#1}}_{R#2}}
\newcommand\Rsq[2]{{\mathcal R}^{\tilde{#1}}_{#2}}

\newcommand\cw[1]{c^{+}_{#1}}

\newcommand\Vmix[1]{V^{}_{#1}}
\newcommand\Umix[1]{U^{}_{#1}}

\newcommand\Zmix[1]{Z^{}_{#1}}

\newcommand\Rsqq[2]{{\mathcal R}^{\tilde{#1} \, *}_{#2}}

\newcommand\cwq[1]{c^{+*}_{#1}}

\newcommand\Vmixq[1]{V^{*}_{#1}}
\newcommand\Umixq[1]{U^{*}_{#1}}

\newcommand\Zmixq[1]{Z^{*}_{#1}}

\newcommand\OwL[1]{\Ocn{L}{#1}}
\newcommand\OwR[1]{\Ocn{R}{#1}}

\newcommand\OwLq[1]{\Ocn{L *}{#1}}
\newcommand\OwRq[1]{\Ocn{R *}{#1}}

\newcommand\auL{\anL{t}{n j}}
\newcommand\auR{\anR{t}{n j}}
\newcommand\buL{\bcL{t}{n i}}
\newcommand\buR{\bcR{t}{n i}}
\newcommand\adL{\anL{b}{m j}}
\newcommand\adR{\anR{b}{m j}}
\newcommand\bdL{\bcL{b}{m i}}
\newcommand\bdR{\bcR{b}{m i}}

\newcommand\auLq{\anLq{t}{n j}}
\newcommand\auRq{\anRq{t}{n j}}
\newcommand\buLq{\bcLq{t}{n i}}
\newcommand\buRq{\bcRq{t}{n i}}
\newcommand\adLq{\anLq{b}{m j}}
\newcommand\adRq{\anRq{b}{m j}}
\newcommand\bdLq{\bcLq{b}{m i}}
\newcommand\bdRq{\bcRq{b}{m i}}

\newcommand\chap{\Charp{i}}
\newcommand\cham{\Charm{i}}
\newcommand\neut{\Neutr{j}}

\newcommand\neutq{\Neutrq{j}}

\newcommand\chape{\tilde{\chi}{}^{+}_{(3-i)}}

\newcommand\sd{\squark{b}{m}}
\newcommand\su{\squark{t}{n}}

\newcommand\sdq{\squarkq{b}{m}}
\newcommand\suq{\squarkq{t}{n}}

\newcommand\OR{\OwR{}}
\newcommand\OL{\OwL{}}
\newcommand\ORq{\OwRq{}}
\newcommand\OLq{\OwLq{}}

\newcommand{\plr}{{\raisebox{0.5pt}
   {$\stackrel{\leftrightarrow}{\partial}$}}{}}

\newcommand{\nn}{\nonumber\\}

\newcommand\buLe{\bcL{t}{n (3-i)}}
\newcommand\buRe{\bcR{t}{n (3-i)}}

\newcommand\Qu{t \,}
\newcommand\Quq{\bar{t}}

\newcommand\Qd{b \,}
\newcommand\Qdq{\bar{b}}

\newcommand\Ruq[1]{\Rsqq{t}{#1}}
\newcommand\Rd[1]{\Rsq{b}{#1}}
\newcommand\Rdq[1]{\Rsqq{b}{#1}}

\newcommand\cona{\tilde{t} t b}
\newcommand\conb{b \tilde{b} \tilde{t}}
\newcommand\conc{\tilde{t} b}

\newcommand{\ownspace}{\vspace*{-10mm}}

\begin{document}
%------------------------------------------------------------------------

\pagestyle{empty} \vspace*{-1cm}
\begin{flushright}
%  draft version \mydate \\
  HEPHY-PUB 804/05 \\
  FI 2005-01\\
  hep-ph/0502112 \\
\end{flushright}

\vspace*{1.4cm}

\begin{center}
\begin{Large} \bf
CP-violating asymmetry in\\[2mm]
chargino decay into neutralino and W boson 
\end{Large}

\vspace{10mm}

{\large H. Eberl$^a$, T. Gajdosik$^b$, W. Majerotto$^a$, and
 B.~Schrau\ss{}er$^{a, c}$}

\vspace{6mm}
\begin{tabular}{l}
$^a${\it Institut f\"ur Hochenergiephysik der \"Osterreichischen
 Akademie der Wissenschaften,}\\
 \hphantom{$^a$}{\it 1050 Vienna, Austria}\\
$^b${\it Institute of Physics, LT-02300, Vilnius, Lithuania}\\
$^c${\it Institut f\"ur Theoretische Physik, TU Graz, 8010 Graz, Austria
% Erzherzog-Johann-University
}
\end{tabular}

\vspace{20mm}

\begin{abstract}
In the MSSM with complex parameters, loop corrections to 
$\tilde\chi_i^\pm \to \tilde\chi_j^0 W^\pm$ lead to a 
CP violating asymmetry $\ACP = 
(\Gamma(\tilde\chi_i^+ \to \tilde\chi_j^0 W^+) -
\Gamma(\tilde\chi_i^- \to \tilde\chi_j^0 W^-))/
(\Gamma(\tilde\chi_i^+ \to \tilde\chi_j^0 W^+) +
\Gamma(\tilde\chi_i^- \to \tilde\chi_j^0 W^-))$. 
We calculate this asymmetry at full one-loop level. We perform a detailed
numerical analysis for $\tilde\chi_1^\pm \to \tilde\chi_1^0 W^\pm$
and $\tilde\chi_2^\pm \to \tilde\chi_1^0 W^\pm$ analyzing the dependence
on the parameters and phases involved. Asymmetries of several percent
are obtained. We also discuss the feasability of measuring these 
asymmtries at LHC.
\end{abstract}
\end{center}

\vfill

\newpage
\pagestyle{plain} \setcounter{page}{2}

It is well known that supersymmetric models contain new sources of
CP violation if the parameters are complex. In the Minimal
Supersymmetric Standard Model (MSSM), the U(1) and SU(2) gaugino
mass parameters $M_1$ and $M_2$, respectively, the higgsino mass
parameter $\mu$, as well as the trilinear couplings $A_f$
(corresponding to a fermion $f$) may be complex. Usually, $M_2$ is
made real by redefining the fields. Non-vanishing phases of $M_1$
and $\mu$ cause CP-violating effects already at tree-level in the
chargino and neutralino production and decay
\cite{ref1,ref2,ref3}. In case the trilinear couplings of the
third generation ($A_t$, $A_b$, $A_\tau$) are complex not only the
stop, sbottom, and stau sectors \cite{ref4} are strongly affected
but also the Higgs sector \cite{ref5,ref6}. The three neutral Higgs bosons
are no more CP eigenstates.
\\
Although new phases in addition to the CKM in the Standard Model
(SM) are desirable to explain baryogenesis, there are severe
constraints on the phase of $\mu$ from the experimental limits on
the electric dipole moments (EDMs) of the electron, neutron and
Hg. For example, in the constraint MSSM $|\phi_\mu|$ has to be
small \cite{ref7,ref8} for a SUSY particle spectrum of the order
of a few TeV.
\\

\begin{figure}[h!]
 \begin{center}
 \mbox{\resizebox{14cm}{!}{\includegraphics{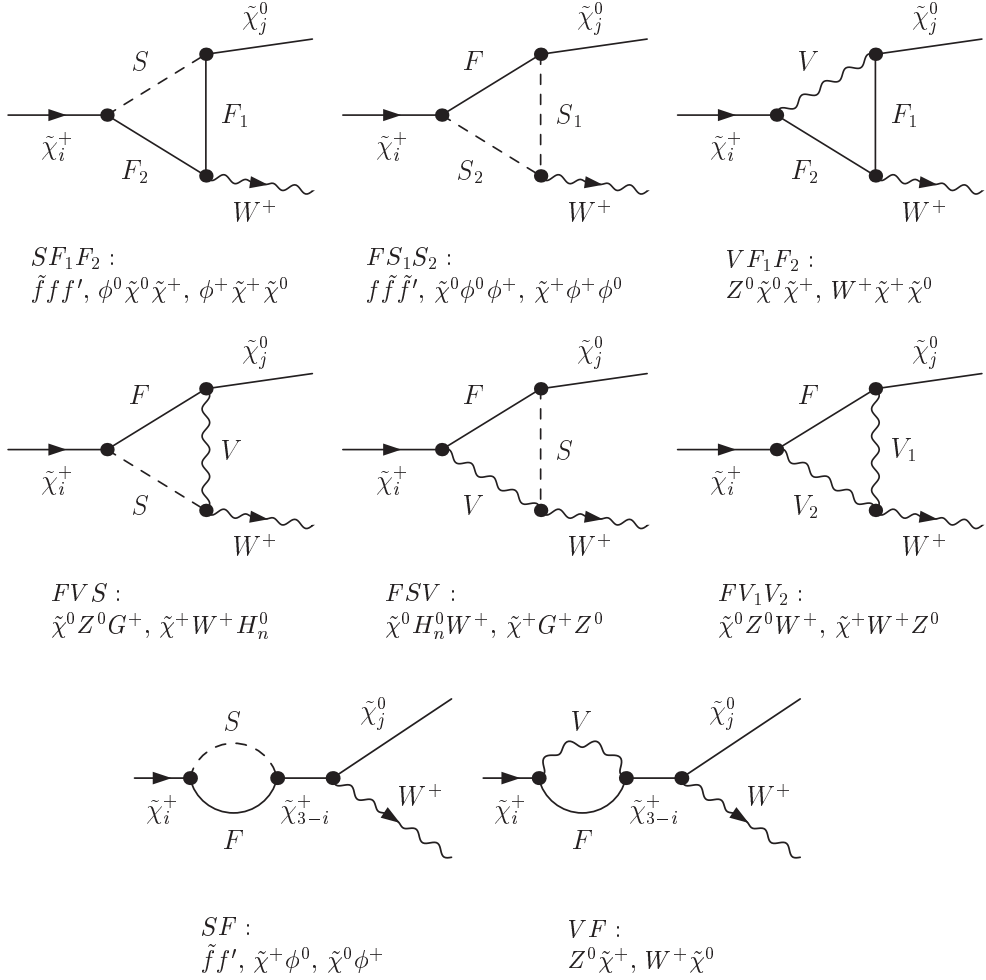}}}
  \end{center}
  \caption[feynman]{All one-loop graphs of the decay
$\chap \to \neut + W^+$, which contribute to the CP asymmetry
$A_{\rm CP}$ defined in eq.~(\ref{ACP}), $f'$ ($\tilde f'$)
denotes the isospin doublet partner of the fermion $f$ (sfermion \sf),
e.g. $f=t$, $f' = b$, $\phi^0 = (H_1^0,H_2^0,H_3^0,G^0)$,
and $\phi^+ = (H^+,G^+)$.
  \label{fig1}}
\end{figure}

In this note, we study CP violation in the decays $\chap (k_2) \to
\neut (k_1) + W^+ (-p)$ and $\cham (-k_2) \to \neut (-k_1) + W^- (p)$ in the MSSM with
complex parameters by calculating the CP-violating asymmetry
\beq
\ACP
=
\frac{\Gamma_{(+)}(\chap \to \neut W^{+})-
      \Gamma_{(-)}(\cham \to \neut W^{-})}
     {\Gamma_{(+)}(\chap \to \neut W^{+})+
      \Gamma_{(-)}(\cham \to \neut W^{-})}
\label{ACP}
\eeq
at full one-loop order. The asymmetry is
%of course
zero if CP is
conserved and also vanishes at tree-level in case of CP violation.
In Fig.~\ref{fig1} we show the graphs which contribute to this
asymmetry at one-loop level. Of course, they give a contribution
to \ACP only if they have an absorptive part, i.e. some
decay channels of $\tilde\chi_i^\pm$ must be open in addition to
that into $\tilde\chi_j^0 W^\pm$.

This asymmetry was already calculated in \cite{ref9} considering
only the third generation quarks and squarks in the vertex graphs.
We have improved this calculation in several points. First, we
performed a full one-loop calculation. In particular, we also
calculated the contributions from self-energies of the charginos.
It turns out, that these are important. (The self-energies of the
neutralinos do not contribute due to their Majorana nature and the
$W^\pm$~-~$H^\pm$ transition vanishes for on-shell $W$-bosons.) In
addition, we take the Yukawa couplings running, which also gives a
sizeable effect. Moreover, we take into account that the neutral
Higgs bosons $(h^0, H^0, A^0)$ mix if the SUSY parameters
mentioned are complex. In our case, this influence is, however,
very small. As a loop-level quantity the asymmetry $A_{\rm CP}$
depends on the phases of all complex parameters involved. One,
however, expects that the dependence on the phases of $M_1$ and
$A_{t,b}$ is strongest (taking $\mu$ real). There is even a strong
correlation between them. Therefore, a measurement of this
asymmetry represents not only a test of CP~violation in
chargino decay, but 
can also be used for the determination of the phases of $M_1$
and $A_{t,b}$.

The widths $\Gamma_{(\pm)}$ can be written as
$\Gamma_{(\pm)} \propto \, | \cMt{(\pm)}|^2 +
  2\Re\big[\cMt{(\pm)\dag} \cMl{(\pm)}\big]$.
 \\
The Feynman amplitudes for the tree- and one-loop level,
$\cMt{(\pm)}$ and $\cMl{(\pm)}$,
are given by
\bea
\cMt{(+)} &=&
i\, \bar{u}_{\neut}(k_1)
    \gamma^\mu ( \OR P_R + \OL P_L )
  u_{\chap}(k_2) \epsilon^*_\mu(-p)
\nonumber \, , \\
\cMt{(-)} &=&
i\, \bar{v}_{\chap}(-k_2)
    \gamma^\mu ( \ORq P_R + \OLq P_L )
  v_{\neut}(-k_1) \epsilon_\mu(p)
\nonumber \, , \\
\cMl{(+)} &=&
i\, \bar{u}_{\neut}(k_1)
  \big[( \gamma^\mu \LpR + k_2^\mu \PpR ) P_R
      +( \gamma^\mu \LpL + k_2^\mu \PpL ) P_L
  \big] u_{\chap}(k_2) \epsilon^*_\mu(-p)
\nonumber \, , \\
\cMl{(-)} &=&
i\, \bar{v}_{\chap}(-k_2)
  \big[( \gamma^\mu \LmR +k_2^\mu\PmL ) P_R
      +( \gamma^\mu \LmL +k_2^\mu\PmR ) P_L
  \big] v_{\neut}(-k_1) \epsilon_\mu(p)
\, . \hspace{6mm}
\eea

Since $|\cMt{(+)}|^2 = |\cMt{(-)}|^2$, and assuming, that the
one-loop contribution is small compared to the tree-level one, the
CP-violating asymmetry \ACP takes the form
\beq
\ACP =
\frac{\Re\big[\cMt{(+)\dag} \cMl{(+)}\big]
     -\Re\big[\cMt{(-)\dag} \cMl{(-)}\big]}
     {|\cMt{}|^2} \, ,
\eeq
with the squared tree-level amplitude
\beq
|\cMt{}|^2 =
  \rho\big(|\OR|^2+|\OL|^2\big)
- 12 m_{\chap} m_{\neut} \Re\big[ \ORq \OL \big]\, ,
\eeq
and the one-loop contributions
\bea & & \hspace{-30pt}
\Re \big[\cMt{(+)\dag} \cMl{(+)} \big]
=
    \rho \Re\big[ \LpR \ORq + \LpL \OLq \big]
- 6 m_{\chap} m_{\neut}
  \Re\big[ \LpR \OLq + \LpL \ORq \big]
\nn & & \hspace{30pt}
+ \frac{\lambda}{2 m_W^2}
  ( m_{\neut} \Re\big[\PpR \ORq + \PpL \OLq \big]
  + m_{\chap} \Re\big[\PpR \OLq + \PpL \ORq \big] )
\, ,
\\[6pt] & & \hspace{-30pt}
\Re \big[\cMt{(-)\dag} \cMl{(-)} \big]
=
    \rho \Re\big[\LmR \OR + \LmL \OL \big]
- 6 m_{\chap} m_{\neut}
  \Re\big[ \LmR \OL + \LmL \OR \big]
\nn & & \hspace{30pt}
+ \frac{\lambda}{2 m_W^2}
  ( m_{\neut} \Re\big[\PmR \OR + \PmL \OL \big]
  + m_{\chap} \Re\big[\PmR \OL + \PmL \OR \big] )
\, ,
\eea
with the kinematic factor $\lambda = \lambda(m^2_{\chap},
m^2_{\neut}, m^2_W)$, $\lambda(x,y,z) = (x-y-z)^2 - 4 y z$,
and $\rho = \frac{\lambda}{m^2_W} + 3 (m^2_{\chap} + m^2_{\neut} - m^2_W)$.\\
The chargino-neutralino-W coupling parameters $O^{L,R}$, defined
by the Lagrangian
\beq
\mathcal{L}_{W \Neutr{j} \Charp{i}} =
  W_{\mu}^{-} \Neutrq{j}
    \gamma^{\mu} ( \OwR{j i} P_{R} + \OwL{j i} P_{L} ) \Charp{i}
+ W_{\mu}^{+} \Charpq{i}
    \gamma^{\mu} ( \OwRq{j i} P_{R} + \OwLq{j i} P_{L} ) \Neutr{j}
\, ,
\eeq
are
\beq \begin{array}{lcl}
\OwR{j i} = g \Zmixq{j2} \Umix{i1} + \frac{g}{\sqrt{2}} \Zmixq{j3} \Umix{i2}
& \mbox{and} &
\OwL{j i} = g \Zmix{j2} \Vmixq{i1} - \frac{g}{\sqrt{2}} \Zmix{j4} \Vmixq{i2}
\end{array}
\, ,
\eeq
where $U$, $V$, and $Z$ are the matrices diagonalizing the chargino and
neutralino system (see eqs. (A.16) and (A.17)).
$\Lambda$ and $\Pi$ are form factors which are given in the Appendix~A.
We only give the form factors for $\tilde\chi^+$
and not for $\tilde\chi^-$, so that $\Lambda, \Pi$ always stands
for $\Lambda_{(+)}, \Pi_{(+)}$. The form factors $\Lambda_{(-)}$
and $\Pi_{(-)}$, belonging to the $\tilde\chi^-$ decay, can be easily
obtained by conjugating all couplings. \\

In Appendix~A we present all formulas for the vertex contributions
with the $\cona$ and $\conb$ loops and the chargino self-energy
contribution with the $\conc$ loop, see graphs \mbox{$S F_1 F_2$},
\mbox{$F S_1 S_2$}, and \mbox{$S F$} of Fig.~\ref{fig1}. The
complete 
analytical formulas will be given in \cite{future work}. All
individual one-loop graphs were numerically checked using the
packages FeynArts, FormCalc, and LoopTools \cite{FeynArts}, 
and FF~\cite{FFpackage}.
We included the CP-violating mixing of the neutral Higgs bosons by
writing our own FeynArts model file. For the numerical program we
used FeynHiggs~\cite{FeynHiggs}.

\section{Numerical results}

We present numerical results for the decay rate asymmetries \ACP
according to eq.~(\ref{ACP}) $\tilde\chi_i^{\pm} \to \tilde\chi_j^0
W^\pm$,  for $i = 1,2$ and $j=1$.
A discussion of the other channels will be given in \cite{future work}. 
For the SM input parameters we take $m_Z$ = 91.1875~GeV, $m_W$
= 80.45 GeV, $\cos\theta_W = m_W/m_Z$, $\alpha(m_Z) = 1/127.9$,
the on-shell parameters $m_t=178$~GeV, and $m_\tau = 1.777$~GeV.
For the bottom mass, our input is the $\MSbar$ value $m_b(m_b) =
4.2$~GeV. For the values of the Yukawa couplings of the third
generation quarks ($h_t$, $h_b$), we take the running ones at the
scale of the decaying particle mass. In principle, the parameters
$A_f$, the U(1) gaugino mass parameter $M_1$ of the neutralino
sector, and $\mu$ can be complex. We
assume that $|M_1| = M_2/2$.
In general, there are 15 independent sfermion breaking mass parameters.
We take $M_{\tilde Q}$ as input and assume the MSUGRA inspired ratios 
$m_{\tilde q} : M_{\tilde Q} : m_{\tilde l} = 3 : 2 : 1$ with
$m_{\tilde q} = M_{\tilde Q_{1,2}} = M_{\tilde U_{1,2}} = M_{\tilde D_{1,2}}$,
$M_{\tilde Q} \equiv M_{\tilde Q_3} =  M_{\tilde U_3} = M_{\tilde D_3}$, and
$m_{\tilde l} = M_{\tilde L_{1,2,3}} = M_{\tilde E_{1,2,3}}$.
In order to reduce the number of input
parameters further, we use $A_t = A_b = A_\tau =: A$. In all figures we take
$M_{A^0} = 300$~GeV, $\tan\beta = 10$, and $\phi_\mu = \pi/10$.

\begin{figure}[h!]
\begin{center}
\mbox{\resizebox{8cm}{!}{\includegraphics{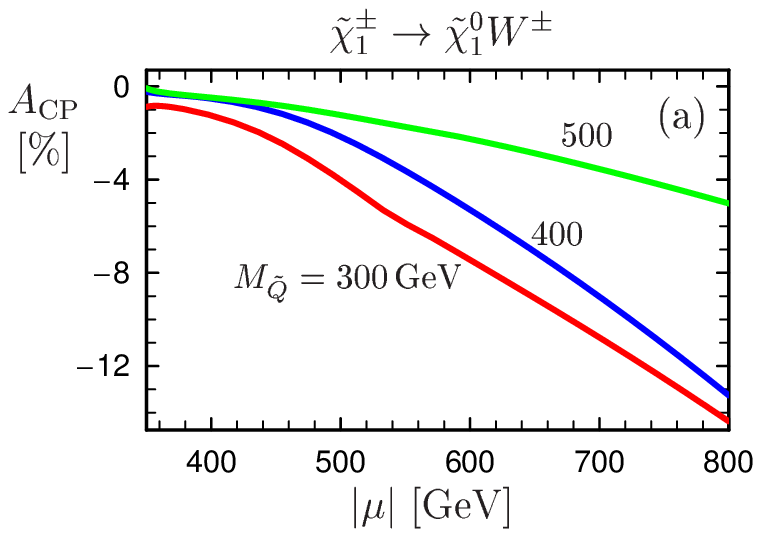}}
\resizebox{8cm}{!}{\includegraphics{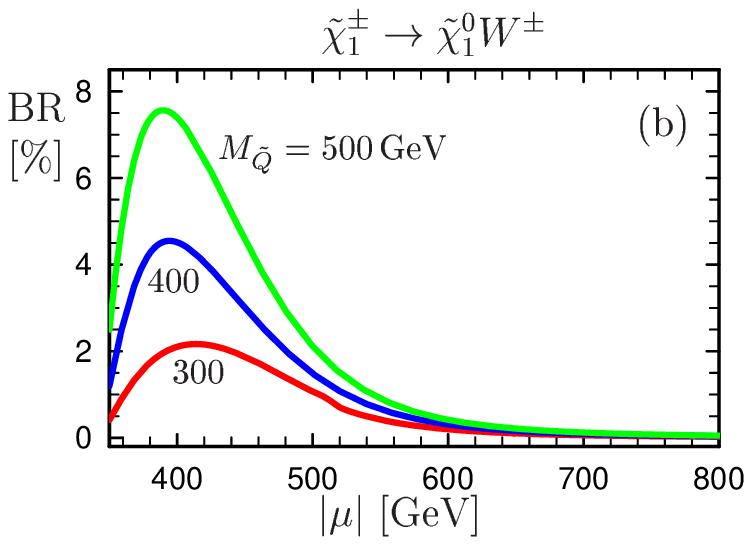}}}
\ownspace
\end{center}
 \caption[fig2]{For
$M_2 = 500$~GeV, $|A| = 400$~GeV, $\phi_A = -\pi/4$, $\phi_{M_1} = 3 \pi/4$,
and three values of $M_{\tilde Q}$, (a) the asymmetry \ACP and (b) the tree-level
branching ratio BR are given as functions of $|\mu|$.
\label{fig2}}
\end{figure}

For the decay $\tilde\chi_1^\pm \to \tilde\chi_1^0 W^\pm$, 
the total one-loop asymmetry \ACP is shown in Fig.~\ref{fig2}a
and the tree-level branching ratio (BR) in Fig.~\ref{fig2}b, for
$M_2 = 500$~GeV, $|A| = 400$~GeV, $\phi_A = - \pi/4$, $\phi_{M_1} = 3\pi/4$,
and three values of $M_{\tilde Q}$ as a function of $|\mu|$.
$|\ACP|$ increases for increasing values of $|\mu|$ because the
tree-level decay width of $\tilde\chi_1^\pm \to \tilde\chi_1^0 W^\pm$ 
goes to zero, as $ \tilde\chi_1^0$ becomes almost a pure bino
which does not couple to $W^\pm$. Therefore, for $|\mu| \gsim 550$~GeV the 
branching ratio drops below 1\%.
The higher the value of $M_{\tilde Q}$ the heavier becomes the stop mass.
Hence \ACP goes down but the branching ratio in (b) increases.

\begin{figure}[h!]
\begin{center}
\mbox{\resizebox{8cm}{!}{\includegraphics{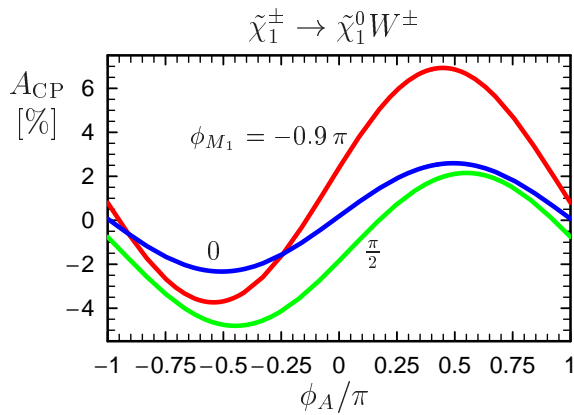}}}
\ownspace
\vspace*{5mm}
\end{center}
 \caption[fig3]{The dependence of \ACP on $\phi_A$, and 
various values of $\phi_{M_1}$, with
$M_2 = 500$~GeV, $|\mu| = 600$~GeV, 
$|A| = 400$~GeV, and $M_{\tilde Q} = 400$~GeV.
\label{fig3}}
\end{figure}

Fig.~\ref{fig3} shows the dependence of \ACP on $\phi_{A}$ for  
$M_2 = 500$~GeV, $|\mu| = 600$~GeV, 
$|A| = 400$~GeV, $M_{\tilde Q} = 400$~GeV,   and various $\phi_{M_1}$.
\ACP has its maximum at $|\phi_{A}| \sim \pi/2$ and is larger at 
large negative values
of the phase  $\phi_{M_1}$. 

\begin{figure}[h!]
\begin{center}
\mbox{\resizebox{8cm}{!}{\includegraphics{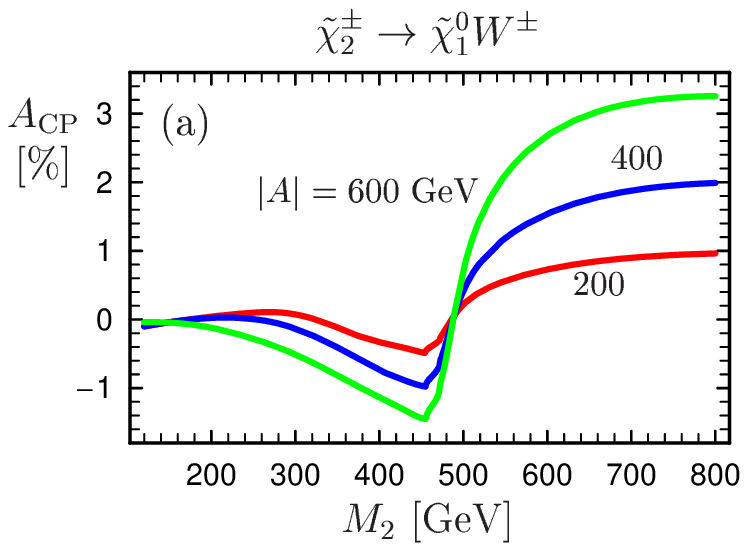}}
\resizebox{8cm}{!}{\includegraphics{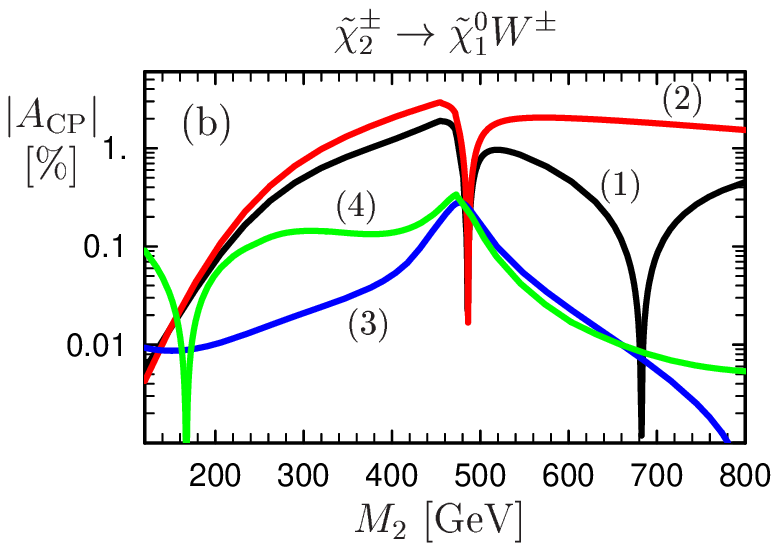}}}
\ownspace
\end{center}
  \caption[plotM2]{The dependence of \ACP  
on $M_2$ for
$\phi_{M_1} = \pi$, $M_{\tilde Q} = 300$~GeV, $\phi_A = -\pi/4$,
$|\mu| = 200$~GeV,

(a) the total asymmetry \ACP for various values of  $|A|$.

(b) the different contributions to the asymmetry
for $|A| = 400$~GeV: the vertex contribution with the
third generation (s)quarks in the loop in black (1); the chargino
self-energy contribution with the third generation (s)quarks in the
loop in red (2); vertex and self-energy corrections with all other
(s)fermions in the loop in blue (3); all remaining corrections in
green (4).
\label{fig4}}
\end{figure}

Now we discuss the asymmetry \ACP for 
$\tilde\chi_2^\pm \to \tilde\chi_1^0 W^\pm$.
Fig.~\ref{fig4}a shows the dependence of the asymmetry $\ACP$ on the 
gaugino mass parameter $M_2$ for various values of $|A|$,
$\phi_{M_1} = \pi$, $M_{\tilde Q} = 300$~GeV, $\phi_A = -\pi/4$,
and $|\mu| = 200$~GeV.  
For $M_2 > 200$~GeV, 
the lighter chargino and the two lighter neutralinos
have dominating higgsino components and the heavier chargino is
mostly gaugino-like ($>90\%$). The bigger $|A|$, the bigger is the mixing in
the squark sector and hence \mbox{\ACP.} Around $M_2 \sim
450$~GeV the $\Charp{2}$ becomes massive enough so that the channels
into $b\tilde{t}_{2}$ and $t\tilde{b}_{1,2}$ open.
For $M_2 \gsim 250$~GeV, the third
generation (s)quark contributions clearly dominate the asymmetry,
the self-energy contribution being bigger than the vertex
contribution. For $M_2 < 680$~GeV, the vertex and the self-energy
contributions for the third generation (s)quarks have opposite
signs and cancel each other to a high degree.
Nevertheless, they remain the dominant contributions 
in a large part of Fig.~\ref{fig4}b.

\begin{figure}[h!]
\begin{center}
\mbox{\resizebox{8cm}{!}{\includegraphics{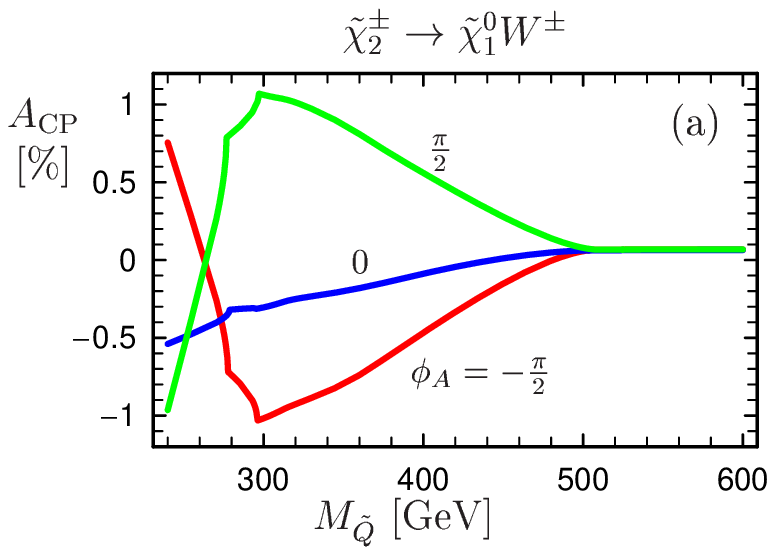}}
\resizebox{8cm}{!}{\includegraphics{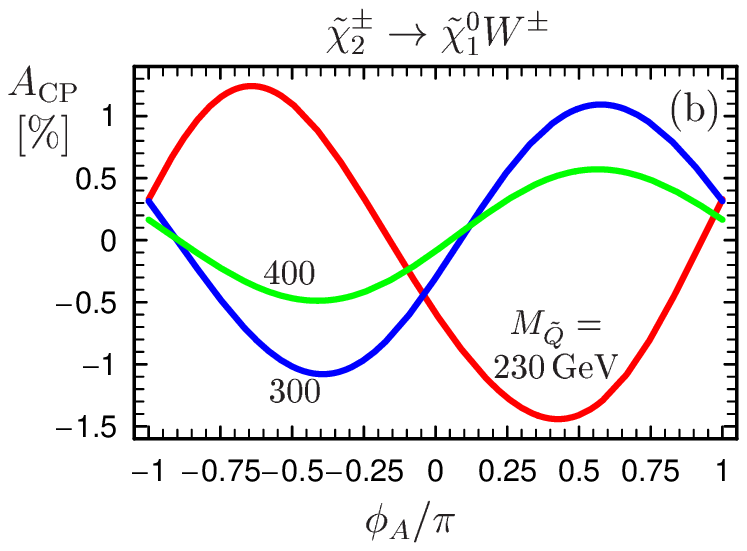}}}
\ownspace
\end{center}
  \caption[plotMSQphiAt]{For $M_2 = 450$~GeV,
$\phi_{M_1} = \pi$, $|A| = 400$~GeV, $|\mu| = 200$~GeV,

(a) the total asymmetry depending on
$M_{\tilde Q}$ for various values of $\phi_A$.

(b) the total asymmetry depending on
$\phi_A$ for various values of  $M_{\tilde Q}$.
  \label{fig5}}
\end{figure}

Various pseudothresholds are visible in Fig.~\ref{fig5}a, where
the squark mass parameter $M_{\tilde{Q}}$ is varied.
The parameter set $M_2 = 450$~GeV, $\phi_{M_1} = \pi$, and $|\mu| =
200$~GeV gives the
masses $m_{\Charp{2}} = 468.55$~GeV and $m_{\Neutr{1}} =
185.66$~GeV.
The strong dependence of \ACP on the phase is clearly visible.
Fig.~\ref{fig5}b illustrates the dependence on the  
phase $\phi_A$  for $M_{\tilde{Q}} =$
\{230, 300, 400\}~GeV. That \ACP does not factorize into a
$\phi_A$ dependent and a $\phi_A$ independent part can be seen
from the fact that the three curves do not meet in a single point.
The other phases $\phi_\mu$ and $\phi_{M_1}$ distort the
factorization.

\begin{figure}[h!]
\begin{center}
\mbox{\resizebox{8cm}{!}{\includegraphics{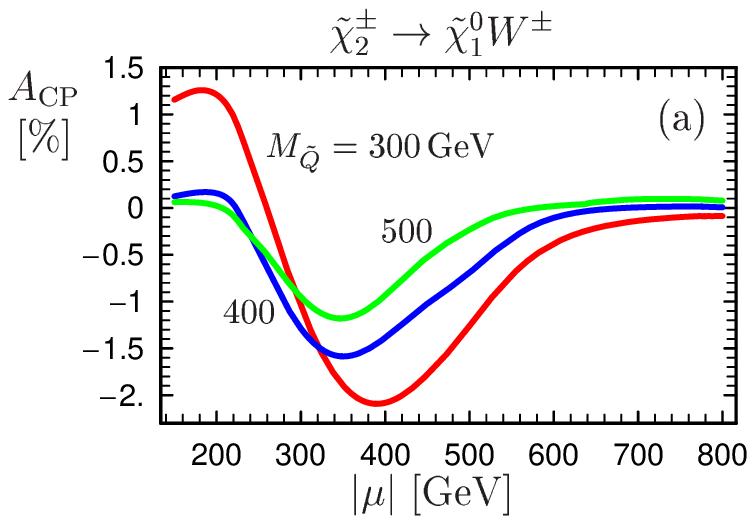}}
\resizebox{8cm}{!}{\includegraphics{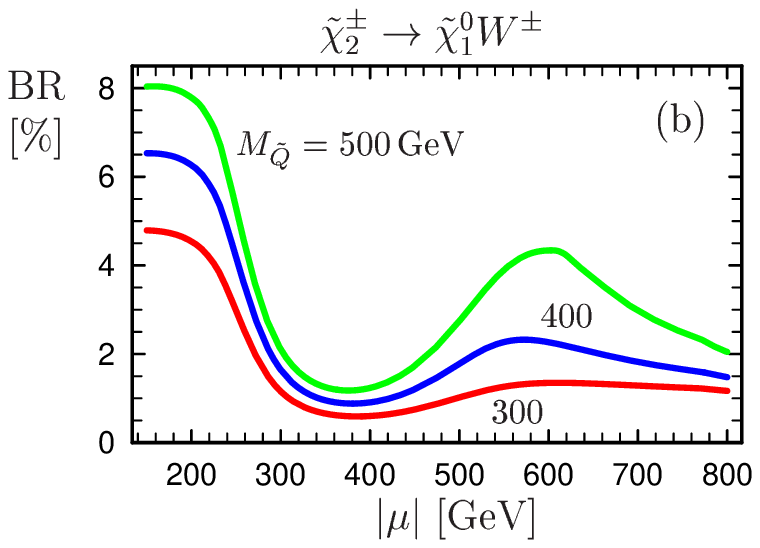}}}
\ownspace
\end{center}
  \caption[plotmue]{For
$M_2 = 500$~GeV, $|A| = 400$~GeV, $\phi_A = -\pi/4$, $\phi_{M_1} = 3 \pi/4$,
and three values of $M_{\tilde Q}$, (a) the asymmetry \ACP and (b) the tree-level
branching ratio BR are given as functions of $|\mu|$.
\label{fig6}}
\end{figure}

The $|\mu|$ dependence of 
the decay $\tilde\chi_2^\pm \to \tilde\chi_1^0 W^\pm$ 
is shown in Fig.~\ref{fig6} 
for the same parameter set as used in Fig.~\ref{fig2}.
The total one-loop asymmetry \ACP is shown in Fig.~\ref{fig6}a
and the tree-level branching ratio (BR) in Fig.~\ref{fig6}b. 
In the region $|\mu| \sim 400$~GeV to 600~GeV the character of 
the $\tilde \chi_2^+$ and $\tilde \chi_1^0$ changes,  
for $\tilde \chi_2^+$ from gaugino to higgsino and for 
$\tilde \chi_1^0$ from higgsino to mainly bino. Therefore, one has a
strong dependence in \ACP and BR there. The dependence on $M_{\tilde Q}$ is 
analogous to that in Fig.~\ref{fig2}. For $|\mu| \gsim 600$~GeV, the mass
of $\tilde \chi_2^+ \sim |\mu|$ and $\tilde \chi_1^0 \sim M_2/2 = 250$~GeV.
Therefore, the decay width of $\tilde\chi_2^\pm \to \tilde\chi_1^0 W^\pm$
increases with $|\mu|$ and \ACP goes to zero. The hump in Fig.~\ref{fig6}b at 
$|\mu| \sim 600$~GeV for $M_{\tilde Q} = 500$~GeV 
is due to the opening of the $\tilde t_2 b$ channel. 

It is known that
the electric dipole moments (EDM) of the electron, the neutron and
mercury strongly depend on the phase of $\mu$ for a light SUSY
spectrum~\cite{Abel:2001vy}. The experimental constraints for the
EDMs of the electron~\cite{Commins:1994gv}, the
neutron~\cite{Harris:1999jx}, and mercury~\cite{Romalis:2000mg}
can be fulfilled by
heavy sfermions of the first generations~\cite{Nath:1991dn}
or if cancellations of different contributions occur \cite{ref8}.
We checked for all plots all three EDMs and found always (small) values
of $\phi_\mu$ that fulfill all EDM constraints.

Finally, we want to comment on the measurability of this asymmetry. At LHC charginos
are mainly produced in the cascade decays of gluinos and squarks so that the production rate
strongly depends on their masses. If the gluino and squark masses are about the same, the gluino
production cross section is far the dominant one. 
With $m_{\tilde g} \sim m_{\tilde q} = 750$~GeV, we expect roughly $2.4 \times
10^5$~events containing $\tilde\chi_1^\pm$ (one has the same amount of $\tilde\chi_1^+$ 
and $\tilde\chi_1^-$ in the case where they
originate from gluinos or from a gluon-gluon process), assuming a luminosity 
of $10^5$~pb$^{-1}$ and a branching ratio of a gluino decaying into a 
$\tilde \chi_1^\pm$ of 40\%. 
Taking into account the branching ratio for 
$\tilde\chi_1^\pm \to \tilde\chi_1^0 W^\pm$, one can measure
\ACP for this decay with a statistical significance of $\sim\,2$
(confidence level of 95\%).
For measuring \ACP for $\tilde\chi_2^\pm \to \tilde\chi_1^0 W^\pm$,
assuming $5 \times 10^4$~events containing a 
$\tilde\chi_2^+$ or $\tilde\chi_2^-$,
one gets a similar statistical significance.

\section{Conclusions}
We have calculated the CP-violating asymmetry between the partial decay
rates $\Gamma(\tilde\chi_i^+ \to \tilde\chi_j^0 W^+)$ and
$\Gamma(\tilde\chi_i^- \to \tilde\chi_j^0 W^-)$ due to phases in the MSSM.
It is a pure loop effect. We have calculated this asymmetry at
full one-loop order. We have given numerical results for
$\tilde\chi_1^\pm \to \tilde\chi_1^0 W^\pm$ and
$\tilde\chi_2^\pm \to \tilde\chi_1^0 W^\pm$. The respective asymmetries
are of the order of several percent, depending on the values of 
parameters and phases involved. In order to have reasonable branching
ratios for the decays the $\tilde\chi_1^0$ must not be very bino like.
We also discussed the feasability of measuring such an asymmetry at LHC.
It might be possible to measured it with a confidence level of 95\%.

\section*{Acknowledgements}
This work was supported by the EU under the HPRN-CT-2000-00149 network 
programme and the ``Fonds
zur F\"orderung der wissenschaftlichen Forschung'' of Austria,
project no.~P13139-PHY. T. G. thanks the Institut f\"ur 
Hochenergiephysik der \"OAW for the hospitality.

\newpage
\begin{appendix}

\section{The (s)top/(s)bottom contributions  to \boldmath{\ACP}}
The relevant parts of the Lagrangian are
% ($\cona$), ($\conb$), and ($\conc$)
\bea
\hspace{-30pt} \LL
&=&
- \frac{g}{\sqrt{2}} \Wm{-} \Qdq \gamma^{\mu} P_{L} \Qu
- \frac{g}{\sqrt{2}} \Wm{+} \Quq \gamma^{\mu} P_{L} \Qd
\\ & &
+ i \cw{m n} \Wm{-} \sdq \plr^{\mu} \su
+ i \cwq{m n} \Wm{+} \suq \plr^{\mu} \sd
\nn & &
+ \Quq \left( \bdR P_{R} + \bdL P_{L} \right) \tilde\chi_i^+ \sd
+ \bar{\tilde\chi}_i^+ \left( \bdRq P_{L} + \bdLq P_{R} \right) \Qu \sdq
\nn & &
+ \Qdq \left( \buR P_{R} + \buL P_{L} \right) \tilde\chi_i^- \su
+ \bar{\tilde\chi}_i^- \left( \buRq P_{L} + \buLq P_{R} \right) \Qd \suq
\nn & &
+ \Quq \left( \auR P_{R} + \auL P_{L} \right) \neut \su
+ \neutq \left( \auRq P_{L} + \auLq P_{R} \right) \Qu \suq
\nn & &
+ \Qdq \left( \adR P_{R} + \adL P_{L} \right) \neut \sd
+ \neutq \left( \adRq P_{L} + \adLq P_{R} \right) \Qd \sdq
\enspace ,
\eea
with the coupling parameters to the $W$~boson,
\beq
\cw{m n} = - \frac{g}{\sqrt{2}} \Rd{m L} \Ruq{n L}
\enspace ,
\eeq
the chargino,
\bea \begin{array}{lcl}
\buR =
- g \Vmix{i1} \Ruq{n L} + g h_{t} \Vmix{i2} \Ruq{n R}
\enspace ,
&&
\buL =
  g h_{b} \Umixq{i2} \Ruq{n L}
\enspace ,
\\[6pt]
\bdR =
- g \Umix{i1} \Rdq{m L} + g h_{b} \Umix{i2} \Rdq{m R}
\enspace ,
&&
\bdL =
  g h_{t} \Vmixq{i2} \Rdq{m L}
\enspace ,
\end{array} \eea
and the neutralino,
\bea \begin{array}{lcl}
\auR =
  g \Ruq{n L} \fL{t}{j} - g h_{t} \Ruq{n R} \Zmix{j4}
\enspace ,
&&
\auL =
- g h_{t} \Ruq{n L} \Zmixq{j4} + g \Ruq{n R} \fR{t}{j}
\enspace ,
\\[6pt]
\adR =
  g \Rdq{m L} \fL{b}{j} - g h_{b} \Rdq{m R} \Zmix{j3}
\enspace ,
&&
\adL =
- g h_{b} \Rdq{m L} \Zmixq{j3} + g \Rdq{m R} \fR{b}{j}
\enspace ,
\end{array} \eea
with the gaugino components of the neutralino
\bea \begin{array}{lcl}
\fR{q}{j} =   \sqrt{2} \charge{q} \tan\theta_{W} \Zmixq{j1}
\enspace ,
&&
\fL{q}{j} =
- \sqrt{2} \left( ( \charge{q} - \isospin{q} ) \tan\theta_{W} \Zmix{j1}
                + \isospin{q} \Zmix{j2} \right)
\enspace ,
\end{array} \eea
and the Yukawa couplings
\bea \begin{array}{lcl}
h_{t} = \frac{m_{t}}{\sqrt{2} m_{W} \sin\beta}
\enspace ,
& \mbox{ and } &
h_{b} = \frac{m_{b}}{\sqrt{2} m_{W} \cos\beta}
\enspace .
\end{array} \eea

The charge and the isospin of the quark $q$ are given by
$\charge{q}$ and $\isospin{q}$, $g$ is the $SU(2)$ coupling
parameter.
\\
The mixing matrices are defined as
\bea
\Umixq{j \alpha} \massmatrix{\charp{}}{\alpha\beta} \Vmixq{k \beta}
&=&
\delta^{}_{jk} m_{\tilde\chi_k^+}^{}
\enspace ,
\\
\Zmixq{j \alpha} \massmatrix{\neutr{}}{\alpha\beta} \Zmixq{k \beta}
&=&
\delta^{}_{jk} m_{\tilde\chi_k^0}^{}
\qquad \mbox{for the basis }
\{ \bino, \wt{3}, \higgsinofield{1}{0}, \higgsinofield{2}{0} \}
\enspace ,
\\
\Rsq{q}{j \alpha} \massQmatrix{\squark{q}{}}{\alpha\beta} \Rsqq{q}{k \beta}
&=&
\delta^{}_{jk} m_{\squark{q}{k}}^{2}
\enspace ,
\eea
with the mass matrices
\bea
  \massmatrix{\tilde\chi^+}{}
&=&
  \left( \begin{array}{cc}
    M_2 &\sqrt 2\, m_W\sin\beta \\
    \sqrt 2\,m_W\cos\beta & \mu
  \end{array}\right)
\enspace ,
\\
  \massmatrix{\tilde\chi^0}{}
&=&
  \left( \begin{array}{cccc}
  M_1 & 0 & -m_Z s_W c_\beta  & m_Z s_W s_\beta \\
  0 & M_2 &  m_Z c_W c_\beta  & -m_Z c_W s_\beta  \\
  -m_Z s_W c_\beta & m_Z c_W c_\beta   & 0 & -\mu \\
   m_Z s_W s_\beta & - m_Z c_W s_\beta & -\mu & 0
  \end{array}\right)
\enspace ,
\\
  \massQmatrix{\sq}{}
&=&
  \left( \begin{array}{cc}  m_{\sq L}^2 & a_q^* m_q \\
                            a_q m_q   & m_{\sq R}^2
  \end{array}\right)
\enspace ,
\eea
where the abbreviations $s_W = \sW$, $c_W = \cW$,
$s_\beta = \sin\beta$, $c_\beta = \cos\beta$, and
\bea
  m_{\sq L}^2 &=&
      M^2_{\tilde{Q}}
    + m_Z^2 \cos 2\beta\,
     ( \isospin{q} - \charge{q} \sin^2 \theta_W )
    + m_q^2
\enspace ,
\\[2mm]
  m_{\sq R}^2 &=&
      M^2_{\sq R}
    + \charge{q} \,m_Z^2 \cos 2\beta\,\sin^2 \theta_W
    + m_q^2
\enspace ,
\\[1mm]
  a_q &=& A_q - \mu^*\, ( \tan\beta )^{-2 \isospin{q}}
          = |a_q|\, e^{i\phi_{\sq}}
\enspace ,
\eea
are introduced for a more convenient notation. $M^2_{\sq R} =
M^2_{\tilde{U}} (M^2_{\tilde{D}})$ is the soft breaking mass
parameter for the right stops (sbottoms).

The form factors giving the major contribution for most of the parameter
regions studied, can be split
into the contributions ($\cona$), ($\conb$), and ($\conc$):
\bea
  \Lambda_{(+)}^{L,R}
& = &\Lambda^{L,R}_{(\cona)}
+ \Lambda^{L,R}_{(\conb)}
+ \Lambda^{L,R}_{(\conc)}
\enspace ,
\\
  \Pi_{(+)}^{L,R}
& = &\Pi^{L,R}_{(\cona)}
+ \Pi^{L,R}_{(\conb)}
\enspace ,
\eea
with
\bea
  \Lambda^{R}_{(\cona)}
&=&
- \frac{3}{16 \pi^{2} \sqrt{2}} \sum_{n=1}^{2} \enspace
    \auR m_{\neut}
    ( \buLq ( C_0 + C_1 ) m_b
    + \buRq ( C_0 + C_1 + C_2 ) m_{\chap} )
\\[-6pt] \nonumber & & \hphantom{-\frac{3}{16 \pi^{2} \sqrt{2}} \sum_{n=1}^{2}}
  + \auL m_t
    ( \buLq C_0 m_b
    + \buRq ( C_0 + C_2 ) m_{\chap} )
\enspace ,
\\[2pt]
  \Lambda^{L}_{(\cona)}
&=&
  \frac{3}{16 \pi^{2} \sqrt{2}} \sum_{n=1}^{2} \enspace
      \auR \buRq ( B_0 - 2 C_{00} + C_0 m_{\su}^{2}
                 + C_1 m_{\neut}^2 + C_2 m_{\chap}^2 )
\\[-6pt] \nonumber & & \hphantom{ \frac{3}{16 \pi^{2} \sqrt{2}} \sum_{n=1}^{2}}
    + \auL \buRq C_1 m_{\neut} m_t
    + \auR \buLq C_2 m_b m_{\chap}
\enspace ,
\\[2pt]
  \Pi^{R}_{(\cona)}
&=&
- \frac{3}{8 \pi^{2} \sqrt{2}} \sum_{n=1}^{2} \enspace
    \auR
    ( \buLq C_2 m_b
    + \buRq ( C_{2} + C_{12} + C_{22} ) m_{\chap} )
\enspace ,
\\
  \Pi^{L}_{(\cona)}
&=&
- \frac{3}{8 \pi^{2} \sqrt{2}} \sum_{n=1}^{2} \enspace
    \buRq
    ( \auR ( C_{1} + C_{11} + C_{12} ) m_{\neut}
    + \auL C_1 m_t )
\enspace ,
\\
  \Lambda^{R}_{(\conb)}
&=&
  \frac{3}{8 \pi^{2}} \sum_{m,n=1}^{2}
    \adL \buLq \cw{m n} C_{00}
\enspace ,
\\
  \Lambda^{L}_{(\conb)}
&=&
  \frac{3}{8 \pi^{2}} \sum_{m,n=1}^{2}
    \adR \buRq \cw{m n} C_{00}
\enspace ,
\\
  \Pi^{R}_{(\conb)}
&=&
  \frac{3}{8 \pi^{2}} \sum_{m,n=1}^{2}
    \cw{m n}
    ( \adR \buRq ( C_{2} + C_{12} + C_{22} ) m_{\chap}
    + \adL \buLq ( C_{1} + C_{11} + C_{12} ) m_{\neut}
\\[-6pt] \nonumber & & \hphantom{ \frac{3}{8 \pi^{2}} \sum_{m,n=1}^{2} c w }
    - \adR \buLq ( C_{0} + C_{1} + C_{2} ) m_b )
\enspace ,
\\[2pt]
  \Pi^{L}_{(\conb)}
&=&
  \frac{3}{8 \pi^{2}} \sum_{m,n=1}^{2}
    \cw{m n}
    ( \adR \buRq ( C_{1} + C_{11} + C_{12} ) m_{\neut}
    + \adL \buLq ( C_{2} + C_{12} + C_{22} ) m_{\chap}
\\[-6pt] \nonumber & & \hphantom{ \frac{3}{8 \pi^{2}} \sum_{m,n=1}^{2} c w}
    - \adL \buRq ( C_{0} + C_{1} + C_{2} ) m_b )
\enspace ,
\\[2pt]
  \Lambda^{R}_{(\conc)}
&=&
- \frac{3}{16 \pi^{2}} \frac{\OwR{j (3-i)}}{m_{\chap}^2 - m_{\chape}^2}
  \sum_{n=1}^{2}
    B_0 m_b ( \buLe \buRq m_{\chap} + \buRe \buLq m_{\chape} )
\\[-6pt] & & \hphantom{ \frac{3}{16 \pi^{2}} \frac{1}{m_{\chap}^2 - m_{\chape}^2} \sum}
  + ( B_0 + B_1 ) m_{\chap}
      ( \buLe \buLq m_{\chap} + \buRe \buRq m_{\chape} )
\enspace ,
\nn
  \Lambda^{L}_{(\conc)}
&=&
- \frac{3}{16 \pi^{2}} \frac{\OwL{j (3-i)}}{m_{\chap}^2 - m_{\chape}^2}
  \sum_{n=1}^{2}
    B_0 m_b ( \buRe \buLq m_{\chap} + \buLe \buRq m_{\chape} )
\\[-6pt] & & \hphantom{ \frac{3}{16 \pi^{2}} \frac{1}{m_{\chap}^2 - m_{\chape}^2} \sum}
  + ( B_0 + B_1 ) m_{\chap}
      ( \buRe \buRq m_{\chap} + \buLe \buLq m_{\chape} )
\enspace \nonumber,
\eea
where
\bea
  C_X^{(\cona)}
&=&
  C_X( m_{\neut}^{2}, m_{W}^{2}, m_{\chap}^{2}
     , m_{\su}^{2}, m_{t}^2, m_{b}^2 )
\enspace ,
\\
  B_0^{(\cona)} &=& B_0( m_{W}^{2}, m_{t}^2, m_{b}^2 )
\enspace ,
\\
  C_X^{(\conb)}
&=&
  C_X( m_{\neut}^{2}, m_{W}^{2}, m_{\chap}^{2}
     , m_{b}^{2}, m_{\sd}^2, m_{\su}^2 )
\enspace ,
\\
  B_X^{(\conc)} &=& B_X( m_{\chap}^{2}, m_{\su}^2, m_{b}^2 )
\enspace .
\eea
The $B$- and $C$-functions are given in the notation of~\cite{denner}.

\end{appendix}

\end{document}